# Skin Cancer Diagnostics with an All-Inclusive Smartphone Application

**Upender Kalwa, Christopher Legner, Taejoon Kong, and Santosh Pandey \***

Department of Electrical and Computer Engineering, Iowa State University,
Ames, IA 50011, USA
**\*** Correspondence: pandey@iastate.edu; Tel.: +1 515-294-7504

**Abstract:** Among the different types of skin cancer, melanoma is considered to be the deadliest and is difficult to treat at advanced stages. Detection of melanoma at earlier stages can lead to reduced mortality rates. Desktop-based computer-aided systems have been developed to assist dermatologists with early diagnosis. However, there is significant interest in developing portable, at-home melanoma diagnostic systems which can assess the risk of cancerous skin lesions. Here, we present a smartphone application that combines image capture capabilities with preprocessing and segmentation to extract the Asymmetry, Border irregularity, Color variegation, and Diameter (ABCD) features of a skin lesion. Using the feature sets, classification of malignancy is achieved through support vector machine classifiers. By using adaptive algorithms in the individual data-processing stages, our approach is made computationally light, user friendly, and reliable in discriminating melanoma cases from benign ones. Images of skin lesions are either captured with the smartphone camera or imported from public datasets. The entire process from image capture to classification runs on an Android smartphone equipped with a detachable 10x lens, and processes an image in less than a second. The overall performance metrics are evaluated on a public database of 200 images with Synthetic Minority Over-sampling Technique (SMOTE) (80% sensitivity, 90% specificity, 88% accuracy, and 0.85 area under curve (AUC)) and without SMOTE (55% sensitivity, 95% specificity, 90% accuracy, and 0.75 AUC). The evaluated performance metrics and computation times are comparable or better than previous methods. This all-inclusive smartphone application is designed to be easy-to-download and easy-to-navigate for the end user, which is imperative for the eventual democratization of such medical diagnostic systems.

## 1. Introduction

Skin is the largest organ in the human body and comprises two distinct layers: epidermis and dermis. While the epidermis protects the body from harsh exposures (such as ultraviolet radiation, infection, injuries, and water loss), the dermis provides nutrition and energy to the epidermis through a network of blood vessels [1–3]. As with every organ in the body, the skin is prone to different forms of cancer. The two most common skin cancers are the basal cell carcinoma and squamous cell carcinoma, which arise from epidermal cells called keratinocytes [4]. A third, deadlier form of skin cancer is malignant melanoma, which develops from epidermal cells called melanocytes. Today, melanoma is notoriously frequent because of increasingly high rates of incidence that lead to a majority of skin cancer deaths [5,6].

To some extent, skin cancer is preventable, and regular screening of skin moles, either in the clinic or at-home, is beneficial for curtailing the progress of the disease. However, current guidelines for screening skin cancer in the United States are inconsistent across different health organizations [7]. For instance, while the American Cancer Society recommends checking for skin cancer during periodic self-examinations by primary care physicians, the American Academy of Dermatology suggests that patients perform skin self-examination without sufficient clarity on the nature and frequency of screening. In a survey involving over 1600 physicians, it was concluded that the most effective skin cancer screening resulted when high-risk patients demanded a complete skin examination and the physicians also had sufficient medical training [7].

As with most cancers, early detection of melanoma can lead to reduced mortality according to several survey studies. In one study, 572 melanoma cases were detected over a 10-year timespan [7]. In another study, 18,000 patients were checked for melanoma over a 17-year timespan. Both studies suggested that the chances of detecting melanoma early on are higher in established patients who routinely visit a skin clinic and are educated on the benefits of routine skin examinations [7]. Besides routine physical examination by a primary care physician or dermatologist, skin self-examination in at-home settings is valuable for the early diagnosis of melanoma. A thorough skin self-examination involves a detailed diagnosis of all body parts, including the back of the body and scalp areas. In addition, imaging technologies aid in accurate diagnosis at an early stage, leading to better treatment and management strategies for melanoma.

The methods to evaluate skin growth for potential prognosis of melanoma have evolved over the past few decades. Before the 1980s, melanomas were generally identified by naked-eye observation of changes in gross mole features, such as large size, bleeding, or ulceration [8]. In the case of suspicious lesions, biopsy of the lesion was done by removing the lesion for further analysis. This invasive method is still the most accurate method for diagnosis of melanoma, but requires the use of trained personnel and expensive equipment. During that time period, early prognosis was difficult because of the lack of technological advancements in imaging hardware and software tools. As time progressed, noninvasive techniques slowly became adopted that entailed less expensive equipment with good accuracy. The most commonly used noninvasive technique is dermoscopy or epiluminescence microscopy, where the skin lesions are examined. A dermoscope is an optical instrument that uses a light source to cancel out skin surface reflections [1,9–12]. This gives access to the in-depth structures and colors of the lesion. This device is connected to a computer and captures images or videos of the lesion that are later used for diagnosis. A sensitivity of 89% has been reported using the dermoscopy method; an improvement over the 70–85% reported for the naked-eye inspection approach [3]. However, the cost is prohibitive for general usage as the dermoscope has to be coupled to expensive equipment (such as stereo microscopes) to evaluate the malignancy of the lesion.

Attempts to democratize skin diagnostics have been demonstrated that use cheaper alternatives to stereo microscopes as an imaging source. A method called named "mobile teledermatology" employed mobile phones to take digital images of the lesion but needed coupling with pocket dermoscopic devices to compensate for the poor-quality optics in early generation mobile devices. The acquired images were transferred to teleconsultants via virtual private networks (VPNs) located at remote locations for analysis and evaluation [13,14]. The two areas of improvement involve: (i) better hardware to capture high-resolution images and (ii) smarter computer-aided diagnosis (CAD) systems to accurately identify melanoma from dermoscopic images. Most of the previously reported CAD systems work on desktop personal computers (PCs) or workstations and assist the physicians to identify cancerous lesions at an early stage so that the treatment regimen can start right away. These CAD systems have generally been tested on dermoscopic or microscopic lesion images, even though they could be integrated with smartphones. Today, mobile phones are equipped with high processing power, more storage capacity, high-resolution image sensors, and larger memory [15,16]. This should enable mobile phones to capture images and run large computational tasks on the image directly on the device itself.

In this work, we developed a smartphone application that functions as an image capture and diagnostic tool for at-home testing of skin cancer. The smartphone optics are enhanced by an inexpensive, commercially available 10x lens. The flow diagram A flowchart of the steps in the entire process involved is illustrated in Figure 1a. A screenshot of the smartphone application classifying a lesion image from a public dataset is shown in Figure 1b. Images of skin lesions are either captured with the smartphone camera or imported from the public dataset. Thereafter, four image-processing steps are implemented: preprocessing, segmentation, feature extraction, and classification. A Gaussian filter reduces the noise, followed by a segmentation algorithm using curve evolution with fast level-set approximations to extract the lesion from the image. A linear affine transformation aligns the lesion axes with the image axes. From the transformed image, the ABCD (Asymmetry, Border irregularity, Color variegation, and Diameter) rule is used to extract features and input them into a support vector machine (SVM) classifier to classify the lesion as benign or melanoma. The development of such all-inclusive skin diagnostic applications is anticipated to gain momentum in coming years, considering the present scenario of health care reforms, expensive costs of hospital visits, and the high mortality rates from melanoma.

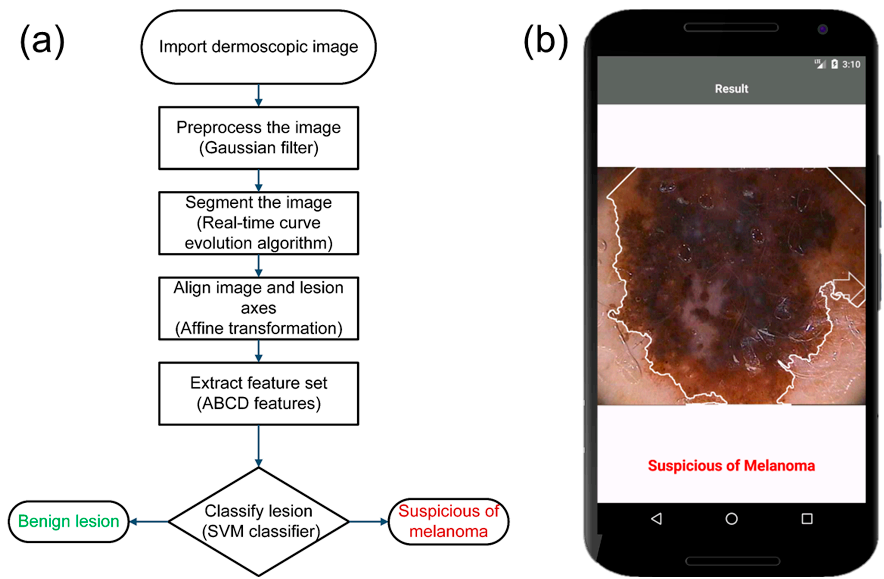

**Figure 1.** Overview of our smartphone application for the prognosis of melanoma. (a) Flowchart shows the different steps involved in the processing of lesion images: The user selects an imported image on the smartphone and selects BEGIN to start the processing stages. The computer-aided diagnosis (CAD) system first preprocesses the image using a Gaussian filter and segments the image using the geometric curve evolution algorithm. The image is rotated to align with the major axes. Asymmetry, Border irregularity, Color variegation, and Diameter (ABCD) features are extracted from the rotated image and classified as benign or melanoma using a support vector machine (SVM) classifier. (b) A screenshot of the final screen of the smartphone application running on the Android Operating System shows a sample lesion classified as melanoma.

## 2. Materials and Methods

Our smartphone application is designed to have relative ease of operation without compromising the accuracy in predicting melanoma cases. The application is intended to have a minimal number of intervention steps from the user with simplistic graphical representation of the classified results. Algorithms should be preferred that can run efficiently on a mobile phone without overloading the computing device. The four stages of image processing performed by the smartphone application are: preprocessing, segmentation, feature extraction, and classification.

*2.1. Preprocessing*

Typically, a dermoscopic image may contain artefacts, such as hairs, ruler markings, air bubbles, and uneven illumination. The first stage towards classifying the malignancy of the skin mole involves preprocessing of the captured image, where the intent is to remove the effects of the abovementioned artefacts, reduce noise, and enhance the image contrast in the image.

In the preprocessing step, color transformation is first performed where the color transformation is first performed, where the RGB image is converted to a different color space that separates the color (chroma) component from the intensity (luma) component. The commonly used color spaces are HSV (where H represents hue, S represents saturation, and V represents value in the cylindrical coordinate system), Y'UV (where Y' represents luminance, and U and V represent the chromaticity components in the Cartesian system), and LAB (where L represents lightness, and A and B represent the chromaticity components in the Cartesian system). We converted the image from the RGB color space to the Y'UV color space, which separates the color component from the illumination component of the image, and is known to perform better in these respects than other color spaces, such as HSV, HIS (where H represents hue, S represents saturation, and I represents intensity in the cylindrical coordinate system), and LAB [17]. This conversion enables the smartphone application to perform consistently under varied illumination conditions (such as indoor and outdoor lighting) and makes the detection of lesion borders less prone to illumination effects. An RGB image is converted to the Y'UV color space based on the National Television System Committee (NTSC) standard using the following equations [18]:

$$Y' = 0.299\,R + 0.587\,G + 0.114\,B \tag{1}$$

$$U = -0.147\,R - 0.289\,G + 0.436\,B \tag{2}$$

$$V = 0.615\,R - 0.515\,G - 0.100\,B. \tag{3}$$

Next, to reduce the effects of artefacts, several filters are available that help to smoothen the image. Commonly used filters are Gaussian [9,19–22], median [16,23,24], and anisotropic diffusion filters [25,26]. We used a two-dimensional (2-D) Gaussian filter $G$ for a point $(x,y)$ represented by Equation (4) below [27]:

$$G(x,y) = \frac{1}{2\pi\sigma^2} e^{-\left(\frac{x^2+y^2}{2\sigma^2}\right)} \tag{4}$$

where $\sigma_x$ and $\sigma_y$ are the standard deviations and $\mu_x$ and $\mu_y$ are the means across both dimensions. The Gaussian filtering produces a resultant image by performing convolution of the filter with the image. The size of the filter is determined by its kernel value. A large kernel value significantly blurs the image and weakens the borderline along with noise, whereas a small kernel value does not reduce the noise to a desirable extent. We found that a kernel of $k = 5$ and a standard deviation of $\sigma = 1$ provided the best results.

*2.2. Segmentation*

In the segmentation step, the lesion boundary is identified from the preprocessed image, which is then used to extract physical features of the lesion. A number of segmentation algorithms have been reported in the literature, such as the edge detection [26,28,29], thresholding [30,31], and active contour methods. previous segmentation algorithms were highly sensitive to noise and thus required high-contrast images in addition to inputs from the user to adjust the segmented region. Today, active contour algorithms have gained popularity, where a deformable curve progresses until it fits the boundary of the region of interest (ROI). Active contour algorithms are categorized as parametric or geometric based on the curve tracking method. In the parametric active contour model, the deformable curve is guided by by energy forces with internal energy controlling the curve's expansion or shrinkage. The image energies (such as image intensities, gradients, edges, and corners) are used to guide the curve to the ROI. Although parametric models have worked even when the ROI has weak borders, there are challenges in handling ROIs with large curvatures and topological changes [10,23]. The geometric active contour model improves upon parametric models by adapting to topological changes. One popular geometric active contour model is known as the Chan–Vese model [25,32]. Generally speaking, active contour models involve solving partial differential equations (PDEs) for curve evolution, creating a computational burden [33].

Our choice of segmentation algorithms was focused around geometric active contour models for the reasons mentioned above, but we desired techniques outside the PDE realm that were computationally light and could run efficiently on a smartphone. We used a modified Chan–Vese model that runs in real-time with fast level-set approximation [34]. In this model, a curve $\phi(x,y)$ denotes a level set function over a grid $u_o$ and is expressed by Equation (6) below:

$$\phi(x,y) = \begin{cases} -3, & \text{if } (x,y) \text{ is an interior point} \\ -1, & \text{if } (x,y) \in L_{in} \\ 1, & \text{if } (x,y) \in L_{out} \\ 3, & \text{if } (x,y) \text{ is an exterior point} \end{cases} \quad (5)$$

where $(x,y)$ represents a point location on the grid. The lists $L_{in}$ and $L_{out}$ contain points inside ($\phi < 0$) and outside ($\phi > 0$) of the curve separated by pixels and allow for localization of the curve. The 'interior points' are the points inside $L_{in}$ and the 'exterior points' are the points outside $L_{out}$. The model updates the curve during each evolution until it fits the boundary of the object of interest. The process of curve evolution is composed of a data-dependent cycle and a smoothing cycle. Both cycles are repeated for $N_a, N_s$ iterations for each evolution. In the data-dependent cycle, a field speed $F_d$ represented by Equation (6) is calculated for all the points in $L_{in}$ and $L_{out}$.

$$F_d(x,y) = \lambda_2(|u_0(x,y) - c_2|^2) - \lambda_1(|u_0(x,y) - c_1|^2) \quad (6)$$

The parameters $\lambda_1$ and $\lambda_2$ are fixed integer values, and $c_1$ and $c_2$ are the mean values inside and outside the curve. The pixel intensity at a point $(x,y)$ on the grid is given by $u_o(x,y)$. For each point in $L_{out}$, if $F_d > 0$, the point is switched from $L_{out}$ to $L_{in}$ and redundant points are deleted. Similarly, for each point in $L_{in}$, if $F_d < 0$, the point is switched from $L_{in}$ to $L_{out}$ and redundant points are deleted. In the smoothing cycle, a speed $F_{int}$ represented by Equation (7) is calculated for all the points in $L_{in}$ and $L_{out}$.

$$F_{int}(x,y) = \begin{cases} 1, & \text{if } G \otimes H(-\phi)(x,y) > \frac{1}{2} \; \forall \, x \in L_{out} \\ -1, & \text{if } G \otimes H(-\phi)(x,y) > \frac{1}{2} \; \forall \, x \in L_{in} \\ 0, & \text{otherwise} \end{cases} \quad (7)$$

where G and H represent the 2-D Gaussian filter and Heaviside functions, and $H(-\phi)$ indicates the object region of the curve ($\phi$). For each point in the $L_{out}$, if $F_{int} > 0$, the point is switched from $L_{out}$ to $L_{in}$ and redundant points are deleted. Similarly, for each point in $L_{in}$, if $F_{int} < 0$, the point is switched from $L_{in}$ to $L_{out}$ and redundant points are deleted. The stopping condition of this segmentation algorithm is defined when the lists $(L_{in}, L_{out})$ are not updated after the first cycle or a set number of iterations is reached. In our case, $\lambda_1 = 2$, $\lambda_2 = 1$ were determined to be the best parameters after multiple trial and error tests.

A cartoon depicting the curve evolution from the initial set curve is shown in Figure 2. In Figure 2a, the lesion image is shown as a 2-D grid with the object of interest depicted in light red. The initial lists $(L_{in}, L_{out})$ are shown by the light green and dark green colors, respectively, where an initially segmented region, based on $\phi$, is overlaid on the original image. During each iteration, the points in two lists are updated in the direction that minimizes the differences in mean values $(c_1, c_2)$. After a certain number of iterations, the curve represented by $C$ fits to the boundary as shown in Figure 2b.

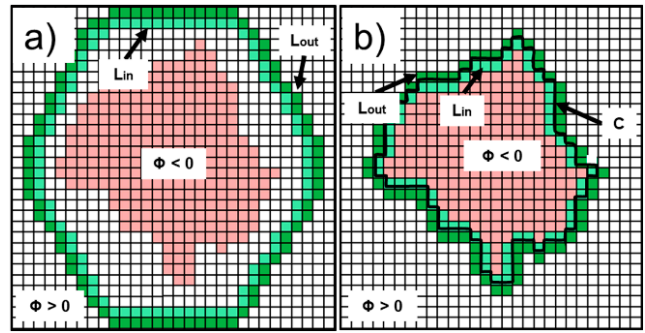

**Figure 2.** Illustration of the geometric active contour curve evolution process during segmentation to identify a lesion. (a) The object of interest is drawn on a two-dimensional (2-D) grid image. The initial elliptic curve with the same height as the image and roughly 80% of the image width along with the initial contours of $L_{in}$ and $L_{out}$ are overlaid on the image. The $L_{in}$, $L_{out}$ contours represent the outer and inner curves, respectively, with a one-pixel gap. All of the points in the image are assigned values based on the level-set function $\phi$. At each iteration, the curves move inward towards the object determined by the speeds. (b) The final contours of $L_{in}$ and $L_{out}$ along with the boundary C of the object after a certain number of iterations are shown.

*2.3. Feature Extraction*

In the 1980s, a group at New York University coined the ABCD acronym to categorize the morphological and clinical features of a skin mole or lesion [35]. This acronym stands for Asymmetry, Border irregularity, Color variegation, and Diameter greater than 6 mm [31,36–38]. The ABCD rule is best suited to differentiate early, thin tumors from benign, pigmented lesions. Besides the ABCD rule, there are other methods and algorithms to detect early melanoma. Pattern analysis has been employed with epifluorescence and video microscopy to categorize the type of skin lesion based upon its general appearance, surface, pigmented patterns, border, and depigmentation [39,40]. Pattern analysis and the ABCD rule are the oldest and most widely adopted methods for melanoma detection [41]. The C.A.S.H. algorithm identifies the Color, Architectural disorder, Symmetry, and Homogeneity/Heterogeneity of mole structures [42]. The C.A.S.H. algorithm has a lower specificity compared to the ABCD rule [8,43]. The Menzies method images the pigmented skin lesions using an immersion oil and categorizes the mole based on the symmetry of pattern and one color [44]. The Glasgow seven-point checklist performs diagnosis on three major features (change in size of lesion, irregular pigmentation, and irregular border) and four minor features (inflammation, itching sensation, diameter greater than 7 mm, and oozing of lesions) [1,3,8,45,46]. Because of its inherent complexity, the Glasgow seven-point checklist is less widely adopted and has a lower pooled sensitivity compared to the ABCD rule [47]. Another method extends the ABCD rule to incorporate the evolution of the lesion (E parameter) by adding the patient's description of lesion change (e.g., enlargement, elevation, and color change) [45,48]. Although the ABCDE rule has been validated in clinical practice, no randomized clinical trials have shown that there is an improvement in the early detection of melanoma [45]. In addition, image acquisition methods have also been developed to differentiate the amount of light absorbed, transmitted, or backscattered by the melanin content of the lesion. Examples of such image acquisition methods are hyperspectral imaging, reflectance confocal microscopy, and optical coherence tomography [43]. However, these image acquisition methods are yet to be standardized to accurately calibrate the absorbance or reflectance from the imaging window [43]. Information about the inflammatory process in the skin has also been retrieved by the use of ultrasound technology and electrical bioimpedance measurements. However, ultrasound images are difficult to interpret and the electrical impedance of the skin can vary greatly based on age, gender, and body location [8]. In addition, advanced dermoscopy and photography tools are commercially available for these applications (e.g., digital epiluminescence (dermoscopic) microscopy (DELM), SIAscope™, MIRROR DermaGraphix™ software, DigitalDerm MoleMapCD™) along with accessory cameras (e.g., Molemax™, SolarScan™, VivaCam™). These equipment are cost prohibitive for at-home diagnostics [49].

In our smartphone application, lesion features are extracted following the ABCD rule. Because of its simplicity in implementation, the ABCD rule is widely adopted and taught in dermatology classes. Furthermore, among all the computerized methods for melanoma detection, the ABCD rule is the most popular and most effective algorithm for ruling out melanoma [47].

The 'Asymmetry' feature stems from the fact that lesion images taken using a dermoscope are generally not symmetric with the major x–y axes of the images. However, to judge if there is any asymmetry in shape, the lesion axes must be aligned to the major axes of the image. To first accomplish this alignment, it is necessary to transform the segmented image ($u_o$) by finding the lesion's minimum enclosing rectangle and extracting the rectangular matrix from the image. This matrix provides the major and minor axes, along with the tilt angle ($\theta$) of the rectangle. Next, we calculate the rotation matrix ($M$) from the tilt angle as shown in Equation (9).

$$\alpha = sf \cos(\theta), \beta = sf \sin(\theta) \tag{8}$$

$$M = \begin{bmatrix} \alpha & \beta & (1-\alpha)C_x - \beta C_y \\ -\beta & \alpha & \beta C_x + (1-\alpha)C_y \end{bmatrix} \tag{9}$$

The parameters $\alpha$ and $\beta$ are calculated from the scale factor and tilt angle as shown in Equation (8). The scale factor ($sf$) determines whether the image should be cropped while rotating or scaling the image so that no information is lost. The application automatically adjusts the scale factor based on the area and position of the lesion in the image. The values ($C_x, C_y$) represent the centroid position of the lesion. From the rotation matrix ($M$), the segmented image is transformed to obtain the rotated image ($R$) as shown in Equation (10) [2]:

$$R(x,y) = u_o(M_{11}x + M_{12}y + M_{13}, M_{21}x + M_{22}y + M_{23}) \tag{10}$$

where $M_{ij} \forall i = 1,2\ j = 1,2,3$ represents the corresponding value at the location ($i,j$) in the rotation matrix.

The asymmetry score is calculated from a total of eight parameters. The first two parameters, vertical and horizontal asymmetry, are calculated by overlapping the binary form of the warped segmented image with the mirror images in horizontal and vertical directions. The sum of all the non-zero pixels in the image is computed and divided by two, assuming that the asymmetrical area will be the same across horizontal and vertical axes. The asymmetry level (AS) is calculated as a percentage of the non-zero pixels in the overlapped image over the lesion area and is represented by Equation (11),

$$AS = \frac{NOR}{A} \times 100 \tag{11}$$

where *NOR* represents the non-overlapped region (non-zero pixels) and *A* represents the lesion area or the total sum of non-zero pixels in the binary image. The remaining six parameters refer to the asymmetry in structure and are calculated as the distance between the lesion centroid and the weighted centroids of the color contours (obtained from the Color variegation feature).

The 'Border irregularity' feature is generally defined as the level of deviation from a perfect circle and measured by the irregularity index (I) as shown in Equation (12),

$$I = \frac{P^2}{4\pi A} \tag{12}$$

where *P* and *A* are the perimeter and the area of the lesion, respectively [31,50]. The minimum value of the irregularity index is the one that corresponds to a perfect circle. As the lesion shape deviates from that of a perfect circle, the value of the irregularity index increases.

The 'Color variegation' feature denotes the different number of colors of the lesion from the HSV (Hue, Saturation, and Value) image. This is calculated by iterating through each pixel of the lesion, extracting its hue value, and grouping all the pixels that have hue values within a specified range. Our color set includes the following colors: white, red, light brown, dark brown, blue-gray, and black.

The HSV values for these colors are determined by trial and error. In general, a benign mole has one or two colors while a melanocytic mole may have more than three colors.

The 'Diameter greater than 6 mm' feature refers to the size of a lesion in the suspicious case of melanoma. However, even with lesions having a diameter of less than 6 mm, the mole should be analyzed for early risks of melanoma. The diameter of the lesion is calculated as shown in Equation (13):

$$D = 2a\gamma \qquad (13)$$

where $a$ is the side length of the minimum area rectangle in pixels, and $\gamma$ is the conversion factor from pixels to millimeters. The $\gamma$ value is calculated using the parameters of the imaging system, such as focal length and the distance from the object to the lens in the system.

*2.4. Classification*

The extracted features are passed to to a classifier that categorizes whether the lesion is suspicious of melanoma or benign in nature. To generate an optimum classifier, the ABCD feature sets from all of the images are randomly divided into training and test sets. Then, a supervised machine learning classification model learns to classify the lesions into different classes based on the input training set. This is generally referred to as the training step. The test set, excluding the classes, is applied on the generated model to create classes which are compared with the ones in the test set to evaluate the model's performance.

The classification algorithm for our smartphone application prioritized robustness, supporting libraries developed for different platforms (Desktop, smartphones) with better performance and faster classification. We chose the SVM classification algorithm because it satisfied all the above conditions and has been demonstrated to work better than other classifiers in other studies [51]. Besides SVM, different classifiers have been implemented for this purpose, including k-nearest neighbor (kNN) [51,52], decision trees [51], and artificial neural networks [53,54] have been implemented for this purpose. The SVM algorithm constructs a discrimination plane (hyperplane) in high-dimensional space that best separates the input data into different classes. Let the training data with $m$ samples be labelled $\{x_i, y_i\}$, where $x_i \in$ input features ($X$), $y_i \in$ input classes ($y$). Let the separating hyperplane constructed by $x$ be defined as $f(x) = w \cdot x + b$, where $w$ is the normal distance to the hyperplane, $|b| / ||w||$ is the perpendicular distance from the origin, and $||w||$ is the Euclidean norm of $w$. The hyperplanes can be formed by $x$ that satisfy $f(x) = 0$ in such a way that the positive samples satisfy $f(x) > 0$ and negative samples satisfy $f(x) < 0$. Depending on the type of data, the following constraints can be formulated [55]:

$$y_i(f(x_i)) - 1 \geq 0 \ \forall \ i = 1, \dots, m \ \text{(Linearly separable)} \qquad (14)$$

$$y_i(f(x_i)) - 1 \geq -\xi_i, \ \xi_i \geq 0 \ \forall \ i = 1, \dots, m \ \text{(Linearly non-separable)}. \qquad (15)$$

For the linearly non-separable data, the variables $\xi_i$, referred to as slack variables, are added such that $\sum_i \xi_i$ sets the upper bound on the total number of errors. An extra cost parameter $C$ is added to assign a penalty for errors. The algorithm chooses the optimum hyperplane based on the largest margin, which is calculated as the sum of the shortest distances from the closest positive and negative sample to the hyperplane. The largest margin is obtained by forming two parallel hyperplanes $H1$ and $H2$. The points that satisfy $\alpha_i > 0$ are called support vectors. These parallel hyperplanes are obtained by minimizing the $||w||^2$ subjected to the inequality constraints (Equation (14) or Equation (15)) depending on the type of data. This minimization is defined by Lagrangian functions for different types of data [55]:

$$L_{ls} = \sum_{i=1}^{m} \alpha_i - \frac{1}{2} \sum_{i,j=1}^{m} \alpha_i y_i \alpha_j y_j x_i \cdot x_j, \qquad \alpha_i \geq 0 \qquad (16)$$

$$L_{lns} = \sum_{i=1}^{m} \alpha_i - \frac{1}{2} \sum_{i,j=1}^{m} \alpha_i y_i \alpha_j y_j x_i \cdot x_j, \qquad 0 \leq \alpha_i \leq C. \qquad (17)$$

The function Equation (17) for linearly separable data ($L_{ls}$) satisfies the conditions $w = \sum_{i=1}^{m} \alpha_i y_i x_i$ and $\sum_{i=1}^{m} \alpha_i y_i = 0$, where $\alpha_i, i = 1, \ldots, m$ are positive Lagrange multipliers for each of the constraints Equation (17) for linearly separable data. Equation (17) for linearly non-separable data ($L_{lns}$) is subjected to conditions $\sum_{i=1}^{m} \alpha_i y_i = 0, i = 1, \ldots, m$.

For nonlinear data, first the samples are mapped to a high-dimensional space ($H$) defined as $\phi: X \to H$. A kernel function, defined as $K(x_i, x_j) = \phi(x_i) \cdot \phi(x_j)$, is used to calculate the dot product of the samples in the higher dimension. The Lagrangian function is modified as [55]:

$$L_{nl} = \sum_{i=1}^{m} \alpha_i - \frac{1}{2} \sum_{i,j=1}^{m} \alpha_i y_i \alpha_j y_j K(x_i, x_j), \qquad 0 \leq \alpha_i \leq C. \tag{18}$$

The above Equation (18) satisfies the conditions $\sum_{i=1}^{m} \alpha_i y_i = 0$, where $\alpha_i, i = 1, \ldots, m$. The kernel function can be set by the user. We tested our smartphone application with three different kernels: Linear, Radial basis function (RBF), and a third-degree polynomial function.

## 3. Results

We initially trained and tested our smartphone application on the publicly available PH2 database [56]. The database has been analyzed by expert dermatologists with added information, such as segmented lesions, identified colors, and their clinical diagnosis. The database consists of 200 dermoscopic images (80 atypical nevi, 80 common nevi, and 40 melanomas) taken by a Mole Analyzer system with a 20x magnification. The RGB images are 8-bit with a resolution of 768 × 560 pixels. To test the images in the database, our smartphone application uses the computer vision library OpenCV for Java. The segmentation algorithm is implemented in C++ and embedded into the smartphone application using Android NDK (native development kit).

Representative images after the preprocessing stage are depicted in Figure 3. In Figure 3a, the first two rows show images of melanoma cases and the last two rows show images of benign cases. We applied the Gaussian filter and tested the dataset with values of the kernel ($k$) ranging from 3 to 11, while maintaining a standard deviation ($\sigma$) value of 1. We found that $k = 5$ and $\sigma = 1$ gave the best results as shown in Figure 3b. A color transformation from the RGB to the YUV color space was performed on the images (Figure 3c).

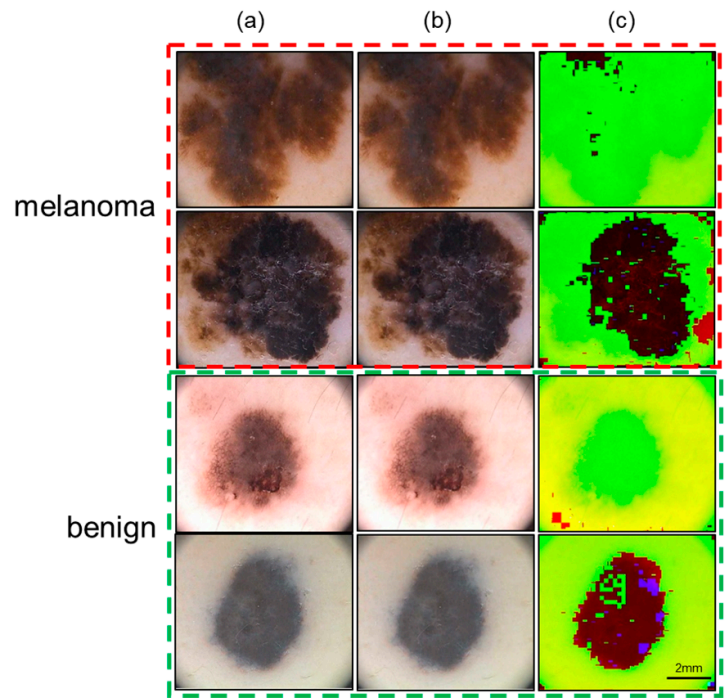

**Figure 3.** The preprocessing stage involves applying a Gaussian filter and color transformation. (a) The column shows the representative images from the publicly available PH2 dataset. The top two

rows are images for melanoma cases and the bottom two rows are images for benign cases. (b) The column shows the results after applying the Gaussian filter with kernel value of 5 and a standard deviation of 1. (c) The column shows the results after converting the color space from RGB to YUV. Scale bar = 2 mm.

Thereafter, the procedure for segmentation is applied to the images in the dataset. The algorithmic parameters $\alpha, \beta,$ and $\gamma$ are used to specify the importance of the Y, U, and V color channels with respect to segmentation and can be adjusted to add more weight to a specific channel, if needed. The initial contour was set at 65% of the image's width and height, and a maximum of 400 iterations were required to find the final contour. Visuals of the curve evolution during segmentation of representative benign and melanoma cases are shown in Figure 4. The first two rows refer to the melanoma cases and the next two rows refer to the benign cases. The curve at different iteration points (50, 100, 200, and 400) is overlaid on the original image in different colors (red, green, cyan, and blue) as shown in Figure 4a, c. It is interesting to observe that the algorithm performs outward evolution for melanoma images and inward evolution for benign images as shown in Figure 4a, c. The final curve is represented by the blue color. The segmented image based on this final curve for these images is shown in Figure 4b, d. If the lambda values $(\lambda_1, \lambda_2)$ are set to values (>2), this seems to affect the segmentation significantly. It is, however, worth noting that, in melanoma cases, the segmentation curve is slightly under-fitted in some cases as shown in Figure 4a, c. In addition, the ABCD values are shown for these images. From the figure, it can be inferred that these values are higher in the melanoma cases.

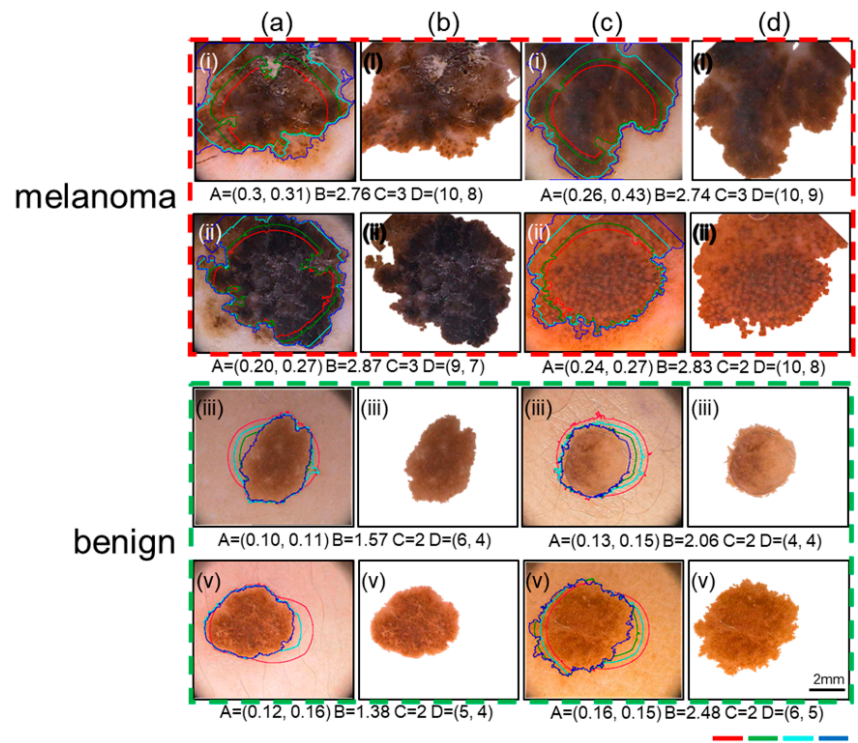

**Figure 4.** The segmentation stage identifies the lesion from the background using the geometric active contour algorithm. The top two rows are images for melanoma cases and the bottom two rows are images for benign cases. Images in the (a) and (c) columns show the original images overlaid with the resulting curve after evolution of 50 iterations (Red), 100 iterations (Green), 200 iterations (Cyan), and 400 iterations (Blue). The images in the (b) and (d) columns show the corresponding final segmented images. The values of the ABCD features are listed. Scale bar = 2 mm.

Our procedure for calculating the asymmetry in shape is depicted in Figure 5 for two representative benign and melanoma cases. The top two rows represent the melanoma cases while the next two rows show the benign cases. Figure 5a, b shows the images after segmenting the lesions from the original images and rotating them to align with the image axes. Figure 5c, d shows the results of the horizontal and vertical asymmetries, in shape, overlaid with their values. The vertical and horizontal asymmetry values, represented by darkened pixels, are higher for melanoma cases than for benign cases.

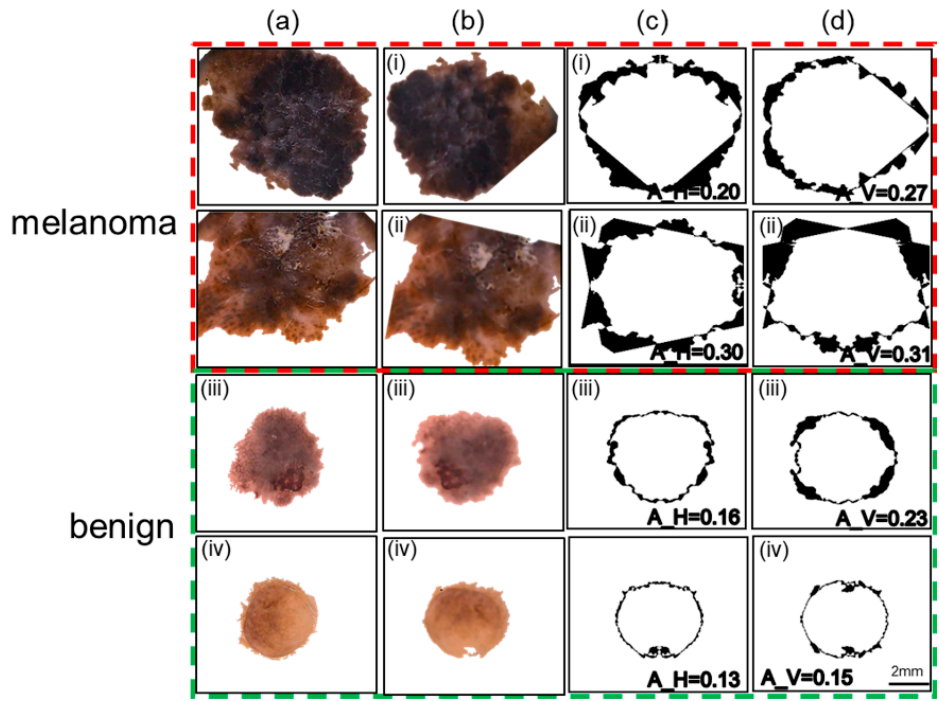

**Figure 5.** The asymmetry in shape is illustrated on representative lesion images. The top two rows are images for melanoma cases and the bottom two rows are images for benign cases. (a) The column represents the images obtained after the segmentation stage. (b) The images are warped (rotated) so that the lesion axes are aligned with the image axes. (c) Horizontal asymmetry is calculated by superimposing the horizontally flipped lesion onto the original lesion and marking the non-overlapped regions with black pixel values. (d) Vertical asymmetry of the lesion is calculated by superimposing the vertically flipped lesion onto the original lesion and marking the non-overlapped regions with black pixel values. The values of horizontal and vertical asymmetry (A_H and A_V, respectively) are listed within the images. Scale bar = 2 mm.

The method to estimate the number of colors (for color variegation) that represents color variegation is applied to the dataset, and representative images (five melanoma and five benign cases) are shown in Figure 6a, c. For most of the benign cases, the number of colors is limited to 2 (light brown and dark brown). However, in the case of melanoma, there are generally more than two colors present. All six color parameters are labelled with different colors: dark brown (red), blue gray (green), light brown (yellow), white (cyan), red (blue), and black (black). The original images with color contours drawn for both melanoma and benign are illustrated in Figure 6b, d.

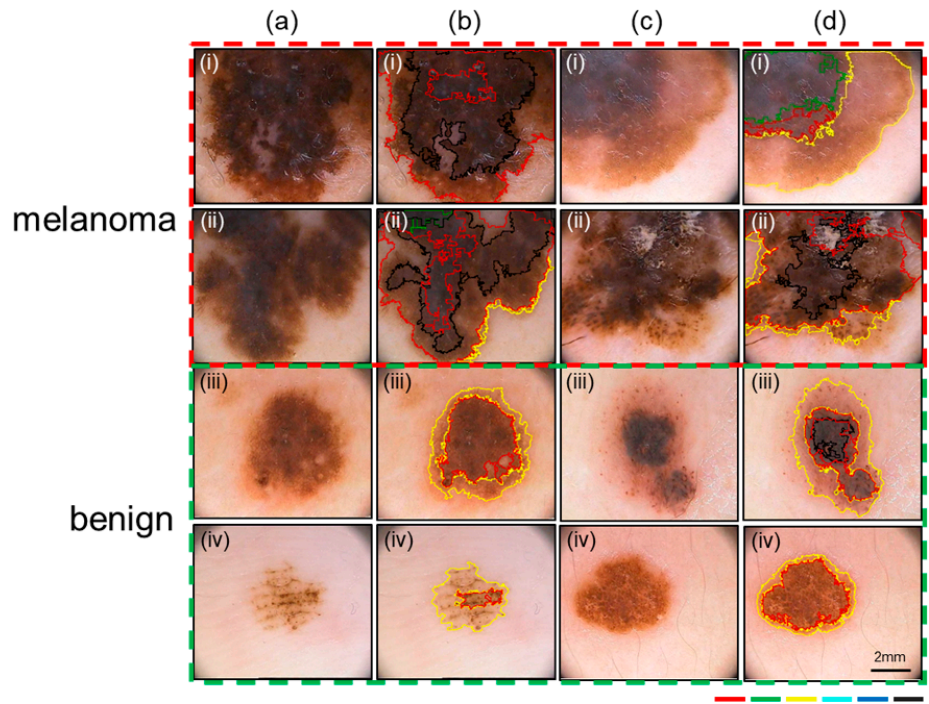

**Figure 6.** The color variegation feature is illustrated showing the different color contours on representative lesion images. The top two rows represent melanoma cases and the bottom two rows represent benign cases. (a, c) The images shown here are the original images from the PH2 dataset. (b, d) The original images are overlaid with the different color region borders detected by the smartphone application. The colors red, green, yellow, cyan, blue, and black correspond to the dark brown, blue gray, light brown, white, red, and black color parameters. Scale bar = 2 mm.

The ABCD features extracted from all of the images are split into training and testing sets with a 70:30 ratio. Because of the low ratio of melanoma to benign cases in the dataset (1:4), we used a popular oversampling algorithm called Synthetic Minority Over-sampling Technique SMOTE [57] to synthetically generate more samples for melanoma to update the training set. The features in the training set are then scaled by subtracting the mean and dividing by the standard deviation for each feature independently maintaining zero mean and unit variance. This scaling is also applied to the testing set using the same mean and standard deviation values from the training set before testing against the classifiers. We evaluated three different kernels for the SVM classifier: Linear, Radial basis function (RBF), and Polynomial against evaluation parameters (sensitivity, specificity, accuracy, and area under curve (AUC)) on the PH2 dataset. We plotted the receiver operating curves (ROCs) and calculated the associated area under the curve (AUC) values (S1). The evaluation metrics for the three kernels are shown in Table 1 for the cases of with and without SMOTE. With SMOTE, the RBF kernel performed better than the other two alternatives (linear and polynomial) for the four evaluation parameters (80% sensitivity, 90% specificity, 88% accuracy, and 0.85 AUC). Without SMOTE, the polynomial kernel performed slightly better than the other two kernels for the evaluation parameters (55% sensitivity, 95% specificity, 90% accuracy, and 0.75 AUC). For studies on early melanoma detection, very high sensitivity is desired. Our calculations indicate that the sensitivity and AUC values with SMOTE were better than those without SMOTE, and the RBF kernel provided the best results with SMOTE.

**Table 1.** Evaluation parameters (i.e., sensitivity, specificity, accuracy, and area under curve (AUC)) were calculated on the PH2 dataset for three SVM kernels (i.e. Linear, Radial basis function (RBF), and Polynomial). The RBF kernel provides the best performance compared to the other two kernels.

| SVM Kernel | With SMOTE | | | | Without SMOTE | | | |
|---|---|---|---|---|---|---|---|---|
| | Sensitivity | Specificity | Accuracy | AUC | Sensitivity | Specificity | Accuracy | AUC |
| Linear | 79 | 82 | 80 | 0.81 | 50 | 92 | 84 | 0.71 |
| Radial basis function (RBF) | 80 | 90 | 88 | 0.85 | 50 | 95 | 86 | 0.72 |
| Polynomial | 79 | 75 | 76 | 0.71 | 55 | 95 | 90 | 0.75 |

The importance of each ABCD feature was evaluated by training the SVM classifier with the RBF kernel. The results are shown in Table 2. When only considering Color variegation, 96% of the melanoma cases were correctly identified, though many false positives were present. Similarly, based on asymmetry alone, 84% of the benign cases were correctly identified, but the increased frequency of false negatives lowered the accuracy and sensitivity values. The diameter feature seemed to have better performance when compared to others. A classifier combining all the ABCD features yielded improved precision (91%) and sensitivity (80%). The associated ROC curves for Table 2 and for multiple SVM kernel parameters are provided in the supplemental file (S1).

**Table 2.** For each ABCD feature, the evaluation parameters (i.e. sensitivity, specificity, accuracy, and precision) were calculated and compared with results from combining all the features on the PH2 dataset.

| Features | Parameters | | | |
|---|---|---|---|---|
| | Sensitivity | Specificity | Accuracy | Precision |
| Asymmetry | 23 | 84 | 54 | 62 |
| Border | 81 | 63 | 72 | 68 |
| Color | 96 | 42 | 69 | 58 |
| Diameter | 90 | 71 | 80 | 75 |
| Overall | 80 | 90 | 88 | 91 |

For all images, a comparison of computational times (in milliseconds) was performed for each processing stage on a desktop (Intel Xeon-E5 CPU, 32 GB RAM, Windows 10) and an Android Phone (Samsung S6). On the desktop, a program was written that runs in Python and makes use of the OpenCV library for Python to perform the image analysis. To speed up the computation time, the same segmentation program used in the smartphone is wrapped using C++ bindings for Python. The results are shown in Table 3. Apart from the segmentation step, the remaining stages take similar computation times for both the benign and melanoma cases on a smartphone and desktop. In both devices, the segmentation step takes longer in melanoma cases (300–400 iterations) when compared to most benign cases (200–250 iterations). The segmentation would normally take longer on a smartphone than on a desktop, but due to the nature of the wrapper call and some additional steps required by the desktop program to extract the lesion boundary, the segmentation duration is longer on the desktop application.

**Table 3.** The computational time of each image-processing stage in our smartphone application is listed as it is run on a Desktop (Intel Xeon-E5 CPU, 32 GB RAM, Windows 10) and Android Phone (Samsung S6) using the PH2 dataset.

| Stage/Device | Benign | | Melanoma | |
| --- | --- | --- | --- | --- |
| | **Desktop PC** | **Android Phone** | **Desktop PC** | **Android Phone** |
| Preprocessing | 78 ± 10 | 116 ± 18 | 68 ± 18 | 109 ± 6 |
| Segmentation | 283 ± 139 | 208 ± 106 | 415 ± 179 | 288 ± 92 |
| Feature Extraction | 27 ± 7 | 41 ± 13 | 31 ± 9 | 47 ± 14 |
| Classification | 10 ± 2 | 19 ± 5 | 10 ± 2 | 19 ± 5 |
| Total Time (ms) | 398 | 384 | 524 | 463 |

The computational time estimated for our smartphone application is lower than those previously reported. For instance, Andrea Pennisi et.al. reported that their system took 1.990 s on an Intel i3-2370M CPU and 4 GB of RAM desktop system to classify an image as melanoma [28]. Aleem et.al. stated that their method took 14.938 s on preprocessed 640 × 480 images using an Android smartphone [22]. Majtner et.al. reported that the computation time to extract the features alone was around 0.6 to 3.3 s [19]. Oliveira et.al. noted that their CAD system took around 8 s per image to classify the lesion on a desktop computer equipped with Intel i5-650 CPU with 8 GB of RAM [25]. Do et.al. reported that their smartphone application classified the lesion, taken by a Samsung Galaxy S4 Zoom smartphone, in less than 5 s after the image is resized with the longer edge at 512 pixels maintaining the same aspect ratio [16]. In comparison, our smartphone application takes less than one second to completely process and classify an image of dimensions 768 × 560 pixels without compromising the accuracy in classification.

After training and testing the smartphone application on images in the PH2 database, we wanted to completely run the developed platform on live images captured using the phone's in-built hardware. The phone is attached with a 10x lens ($11, AMIR camera lens kit, Shenzhen Amier Technology) to allow us to take microscopic images of the skin moles. Eight individuals, each having different skin tones, volunteered to have images of their moles captured by a smartphone (Figure 7a). The operational procedure is as follows. First the 'Mole Detection' application is opened, and the 'Begin' button is pressed providing the user with two features ('Take a photo' and 'Get stored pictures'). When the 'Take a Photo' feature is chosen, the application takes control of the smartphone's camera, allowing it to capture an image at the user's discretion. The user points the smartphone camera towards the lesion and captures an image after it is focused. This happens when the lens is roughly 13 mm away from the target lesion. This value is also used to recalculate the pixel to a mm conversion factor and update it in the application. While capturing an image, a support structure allows us to align the smartphone perpendicularly to the lesion at the optimal focal distance (Figure 7a(i)). An example L-shaped glass support structure (L = 75 mm, W = 75 mm, H = 13 mm) is shown in Figure 7a(ii) with a sample mole image taken using this support structure (Figure 7a(iii)). We have seen that focused images (focal distance = 13 mm) captured under white light or indirect sunlight are adequately processed by the smartphone application. Once the image is captured, it is resized to 1024 × 768 using bilinear interpolation before storing it. The user is then provided with two options ('Retake' or 'Proceed'). If the user is unsatisfied with the image quality, they can retake a new image by selecting the 'Retake' option. Otherwise, the user can select the 'Proceed' option to perform the diagnosis of the lesion. The user is presented with the final classified results (i.e., benign or suspicious of melanoma) along with values of the ABCD features. Figure 7b shows the original mole images overlaid with segmentation contours. It can be observed that the segmentation results are more sensitive to borders, showing more irregularity than appears visually. This may be due to the high resolution (2988 × 5312 pixels) of the S6 camera, which is not improved after resizing to 1024 × 576 when using bilinear interpolation and a smoothing filter. Even with this drawback, all images were successfully processed and classified as benign. A supplemental file (S2) demonstrates the usage of our smartphone application on both the PH2 dataset and live images. In addition, we have tested our system on a digital lesion images MED-NODE public dataset [5] and our sensitivity, selectivity, and accuracy values are 70%, 80%, and 75%, respectively. Some of the sample-processed mole images

from this dataset are included in the supplemental file (S1). The software is available at *https://github.com/ukalwa/melanoma-detection* for the interested readers who want to build upon our current prototype.

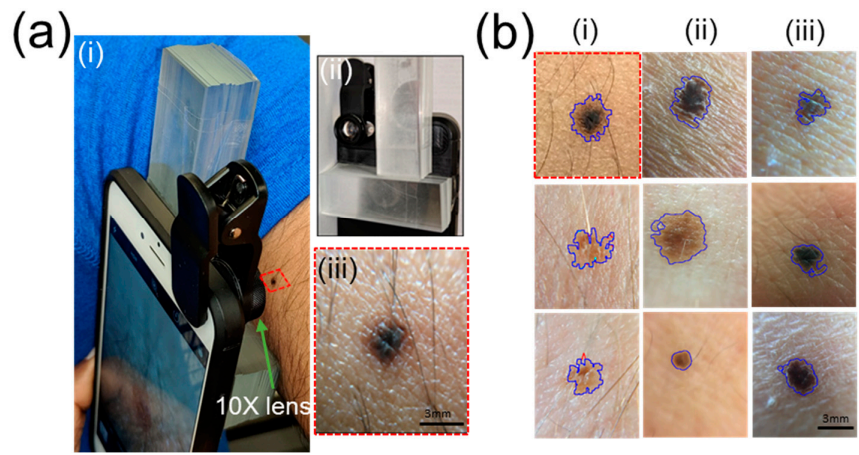

**Figure 7.** The smartphone application is used to capture, process, and classify live images of skin lesions. (a) The picture shows the setup that includes a support structure (L = 75 mm, W = 75 mm, H = 13 mm) to capture a person's mole with our application (i). The glass support structure (ii) helps to align the camera onto the mole to capture the image at the appropriate focal distance (iii). A Macro 10x lens is attached to a smartphone rear camera and focused on a mole on the hand. Our application captures the image and processes it to determine the malignancy of the mole. (b) Representative results for moles from eight volunteers are shown here overlaid with segmentation contours. All the moles were classified as benign here. Scale bar = 3 mm.

## 4. Discussion

Looking at the reported melanoma cases over the past few years, there are some insightful points that can be enumerated. According to the statistics released by the American Cancer Society for the year 2018, there are 91,270 new cases (55,150 males; 36,120 females) of melanoma in the United States. Out of this population, there have been an estimated 9320 deaths (5990 males; 3330 females). The states with the highest number of new melanoma cases in 2018 are California (9830), Florida (7940), New York (4920), Texas (4440), and Pennsylvania (4320) [58]. In comparison to the statistics from 2014, there were 76,100 new cases (43,890 males; 32,210 females) of melanoma within the United States, with an estimated 9710 deaths (6470 males; 3240 females) [59]. The above data indicate that, in the United States, both the number of new melanoma cases and fatalities have increased in the past five years. Furthermore, the states reporting a higher incidence of melanoma correspond with sunny regions and larger populations, where it is challenging to screen and educate the masses about skin cancer and early diagnosis. It can reasonably be assumed that the statistics on melanoma incidence and fatalities would be worse in the developing countries where access to diagnostic/preventive medical resources is still sought. More importantly, the statistics indicate that there is still an urgent need for portable melanoma screening devices that can be readily adopted.

Smartphone-based skin cancer recognition remains a challenging area of research, and this has slowed the commercialization and general availability of portable melanoma screening devices. Some of the technological limitations of the previously reported methods include: (a) the application used a non-smartphone-based camera with a dermatoscope to photograph lesions [13,14,19,20,29], (b) the application was tested on small image set or different databases, making it difficult to directly compare the results [15,16,19,22], (c) the application demonstrated a single feature extraction [2,30,60], (d) the application ran only on a desktop and thus was not considered real-time [13,19,21,25,54], and (e) the application had average accuracy, sensitivity, or selectivity [22,50,51]. We showed that our smartphone application is able to overcome the above limitations with better or comparable computation times and accuracies to those reported earlier.

The scope of our work can be extended in terms of performance and usage. The performance parameters (i.e., accuracy, sensitivity, selectivity, and processing time) could be improved by using better thresholding algorithms for boundary detection, testing other color space parameters (shape, color, and texture distributions), and running it on a large variety of images. The application runs solely on a smartphone, so its usage could entail tracking the progression of specific moles over a period of time. To do this, the user would capture and save mole images at different time intervals over a two-week period, recording the ABCD parameter values for each image. From these recorded values, it would be possible to observe any changes in the shape, size, or color of the mole that may then be evaluated by a healthcare professional. The application's reliability and reproducibility can be tested on persons with different skin colors, under different background illumination conditions, and with different stages of lesion malignancy. With the sophistication of smartphone optics today, it may be possible to directly visualize a lesion's distribution in the different skin layers leading to accurate identification of lesion attributes that may be missed, especially during the early stages of melanoma. Due to the challenge of forming collaborative relationships with skin clinics, most researchers have relied on public databases for their research. This makes direct comparisons of performance metrics across different methods difficult. The present study can be expanded beyond the ABCD rule to understand the role of each feature in the eventual classification of melanoma or benign lesions. The commonly used discriminators include shape features (e.g., asymmetry, aspect ratio, maximum diameter), color features in the color spaces (mean distance, variance, maximum distance), and texture features (e.g., grayscale co-occurrence matrix and texture descriptors). While all three feature types are equally relevant in accurate discrimination and classification, the color features have been shown to perform better than the texture features [21].

## 5. Conclusions

The current prototype comprises a smartphone application to capture or import images of skin lesions, perform feature extraction based on the ABCD rule, and classify their malignancy based on the SVM classifier. The application was tested on 200 dermoscopic images from the PH2 database and the benign moles of two individuals. The entire process from image capture to classification runs entirely on an Android smartphone equipped with a detachable 10x lens, and has a processing time within one second per image. Easy-to-use navigation buttons are incorporated at the front-end to assist the user through the various processing steps. For the PH2 database, the overall performance is better with SMOTE (80% sensitivity, 90% specificity, 88% accuracy, and 0.85 AUC) compared to without using SMOTE (55% sensitivity, 95% specificity, 90% accuracy, and 0.75 AUC). Scope for improvement lies in training with even larger image datasets, having access to individuals with possible melanoma cases, and testing under varied environmental conditions and disease stages.

**Conflicts of Interest:** No conflicts of interest exist in this work.

**Acknowledgments:** The authors are grateful to the image database (PH2 and MED-NODE) for providing public access to the images. This work is partially supported by the U.S. National Science Foundation [NSF IDBR-1556370] and Defense Threat Reduction Agency [HDTRA1-15-1-0053] to S. P.